\documentstyle[12pt,epsf]{article}

\newcommand{\beq}{\begin{equation}}
\newcommand{\eeq}{\end{equation}}
\newcommand{\bea}{\begin{eqnarray}}
\newcommand{\eea}{\end{eqnarray}}

\newcommand{\nbox}{{\,\lower0.9pt\vbox{\hrule \hbox{\vrule height 0.2 cm \hskip
0.2 cm \vrule height 0.2 cm}\hrule}\,}}

\DeclareFixedFont{\xiiss}{OT1}{cmss}{m}{n}{12}
\DeclareFixedFont{\ixss}{OT1}{cmss}{m}{n}{9}
\DeclareFixedFont{\cmrnine}{OT1}{cmr}{m}{n}{9}

\newcommand{\CC}{\hbox{\xiiss C\kern-.4emI}}
\newcommand{\RR}{\hbox{\xiiss R\kern-.45emI}}
\newcommand{\ZZ}{\hbox{\xiiss Z\kern-.4emZ}}
\newcommand{\CCs}{\hbox{\ixss C\kern-.4emI}}
\newcommand{\ZZs}{\hbox{\ixss Z\kern-.4emZ}}
\newcommand{\pa}{\partial}

\newcommand{\pasl}{\pa\kern-.55em /}
\def\href#1#2{#2}

\begin{document}
\begin{titlepage}
\title{ 
        \begin{flushright}
        \begin{small}
        RU-NHETC-2000-16\\
	PUPT-1928 \\
        hep-th/0005047\\
        \end{small}
        \end{flushright}
        \vspace{1.cm}
Near Hagedorn Dynamics of NS Fivebranes \\
{\it or }\\
A New Universality Class of Coiled Strings\\}

\author{
\\
\\
Micha Berkooz\thanks{e-mail: \tt mberkooz@feynman.princeton.edu} \\
\small \it Department of Physics\\
\small  \it Princeton University\\
\small \it Princeton, NJ 08544 \\
\\
\and
\\
\\
Moshe Rozali\thanks{e-mail: \tt rozali@physics.rutgers.edu}\\
        \small\it Department of Physics and Astronomy\\
        \small\it Rutgers University\\
        \small\it Piscataway, NJ 08855
}

\maketitle
\begin{abstract}

We analyze the thermodynamics of NS 5-branes as the temperature
approaches the NS 5-branes' Hagedorn temperature, and conclude that
the dynamics of ``Little String Theory'' is a new universality class
of interacting strings. First we point out how to vary the temperature
of the near extremal solution by taking into account $g_s$
corrections. The Hagedorn temperature is shown to be a limiting
temperature for the theory. We then compare the thermodynamics to that
of a toy model made of free strings and find basic discrepancies. This
suggests a need for a new class of string interactions. We suggest
that this new universality class is characterized by a strong
attractive self-intersection interaction, which causes strings to be
coiled. This model might also explain why ``Little String Theories''
exist in at most 5+1 dimensions.

\end{abstract}

\end{titlepage}

\section{Introduction}

One of the more puzzling objects in string theory is the NS
fivebrane. In particular the worldvolume theory, the so-called
``Little String Theory'' (LST), is believed to be some kind of
non-gravitational string theory.  These theories were introduced in
\cite{brs,seiberg} (see also \cite{dvva}, for a review see
\cite{ofer}). In particular in \cite{seiberg} the decoupled theory on
a cluster of NS fivebranes was defined by taking the limit
\beq
g_s\rightarrow 0,\ \ \  M_s\ fixed
\eeq
of string theory in the presence of the cluster. This definition,
however, is rather indirect in the sense that it does not provide a
microscopic description of the theory. Rather, it is closely related
to the gravitational holographic dual \cite{juan} (for a review see 
\cite{jrev}) of the theory.

The gravitational side of the holographic duality for this case
\cite{abks} is the throat region of the CHS background
\cite{chs}. This background includes a linear dilaton direction, an
$R^6$ component and a WZW model at a level set by the number of
fivebranes $N$. This duality was used to analyze the observable
content of the theory at the origin of its moduli space \cite{abks},
and along its flat directions \cite{gk}. This was done both for the
type II NS fivebranes, the heterotic NS fivebranes \cite{het},
fivebranes at orbifolds \cite{other} and lower dimensional related
configurations \cite{gkp}. Fivebrane thermodynamics was discussed in \cite{emil}.

A complementary way of exploring ``little string theories'' was
introduced in \cite{abkss,higgs} and developed further in
\cite{dlcq,ab}. In this approach a discrete light cone quantization
\cite{lenny} was suggested, along the lines of Matrix theory
\cite{matrix}. This DLCQ description is in terms of a 1+1 sigma model
on the ADHM moduli space. Even though some aspects of this model are
well understood (either from AdS/CFT or from a direct field theory
analysis \cite{ab}), it is not yet clear how to directly obtain LST
quantities from it and in particular how to obtain a covariant
microscopic description (we suggest a way of going to the ``long
strings'' picture for this sigma-model in section 6). For another suggestion,
involving quasi-local field theories, see \cite{anton}.

Despite the difficulties, it is worthwhile to explore these theories
for a variety of reasons. The basic point is that these are stringy
theories without gravity and without a tunable string coupling. This
makes them a very intriguing object to study. In addition \cite{brs},
they are important for a Matrix description of M-theory on
$T^5$. Moreover, in the context of holographic duality the ``little
string theory''/CHS background duality behaves differently from the
AdS/CFT duality, and therefore might be a first step towards
understanding holography in asymptotically flat spaces \cite{abks}.

Another interesting aspect of these theories is an unusual UV/IR
relation. When the theories are taken along the flat directions
\cite{gk, private}, such that the Higgs vev is $M_w^2$, then in
addition to the expected massless fields ($N$ tensor(vector)
multiplets for type IIA(IIB)), one finds additional massless
states. In fact one finds an entire stringy tower of massive states
with string scale $M_s/\sqrt{N}$, even in the limit $M_w\rightarrow
\infty$. This is very puzzling because naively one expects the theory
in this limit to split into $N$ copies of the theory of a single NS
fivebrane. Rather the low energy theory is still sensitive to details
of the very high energy states. This is somewhat reminiscent of
effects in other non-local quantum theories, i.e., the theories on
non-commutative geometries \cite{msv}. The difference is that the
extra states here are propagating particles in Minkowski space and the
theory is Lorentz invariant. 

More recently, these theories have appeared in the context of
holographic duals of confining gauge theories \cite{joemat}. Some of
the confining vacua of ${\cal N}=4$ deformed to ${\cal N}=1$ pure glue
are described holographically using NS fivebranes. This suggests a
role for ``little strings'' in the confinement picture, or more generally
as an approximate description, valid at some finite energy interval,
of the IR region of field theories.

In this paper we discuss the behavior of LST at high energy densities.
The holographic dual to this configuration is the near horizon limit
of the near extremal fivebranes \cite{horstro,gidstro}, which includes
the CGHS black hole \cite{witten,cghs}. We are interested in the
Euclidean black hole which means that we are discussing the canonical
ensemble. From this background it is easy to extract that the theory
has an Hagedorn density of states and hence a Hagedorn temperature
(for discussions see \cite{ofertom,gk, dbranes}). We are interested at
the behavior of the theory as the temperature approaches the Hagedorn
temperature from below.

The behavior at high energies is sometimes expected to resemble
weakly coupled string theory \cite{juan1, ofertom}.  We attempt to
interpret the thermodynamics in terms of a gas of weakly coupled
strings. To this end we begin, in the next section, by reviewing a
class of models of free strings and the resulting thermodynamics. This
is meant to be compared to qualitative features, and not necessarily
precise quantitative ones, of the LST thermodynamics. We discuss
possible different behaviors near the Hagedorn temperature, the
validity of the canonical ensemble and other features.

In sections 3 and 4 we discuss the thermodynamics of the CGHS black
hole. Section 3 is a review of known results at string tree level and
section 4 discusses one loop corrections, which allow us to go off the
Hagedorn temperature and study the partition function as we approach
it. The results strongly suggest that the Hagedorn temperature is a
limiting temperature for LST, rather than a  phase transition
as in  weakly coupled critical  string theory. 

Section 5 is somewhat of an aside - we discuss how the large
fluctuations of the canonical ensemble manifest themselves in the near
extremal solution.

In section 6 we try to interpret our results in terms of the dynamics
of an almost free string and show that there is some qualitative
difference between the two systems. We interpret this as indicating a
new universality class of interacting strings. We then suggest that in
this universality class the strings have a strong self-attractive
potential. This makes long strings want to shrink, which is a first
step towards explaining the thermodynamics of the NS 5-branes. By a
simple combinatoric random walk model we argue that this phase can not
occur for strings above 5+1 dimensions, which explains the maximal
dimensionality of ``little string theory'' (although it does point to
the fact that a similar modification for the theory of membranes might
lead to an interacting theory even in higher dimensional spacetimes). Our
analysis is based both on space-time considerations and on DLCQ
considerations, and points to a new way of analyzing ``long strings''
in the D1-D5 system.

We conclude by summarizing the main results of the paper.

As we were completing this project, we received a paper \cite{obers}
with some overlap with sections 3 and 4 in our paper. The
interpretation of the result is, however, quite different.

\section{Ensembles of Weakly Coupled Strings}

Let us review some aspects of the thermodynamics of critical string
theory, which are relevant for us (these would be similar to
\cite{aw,strings}.). Of course, the notion of a thermodynamic
equilibrium in a theory with gravity is ill defined, but as explained
in \cite{aw} it is justified in the weak string coupling limit, where
some questions can still be addressed. The question we are interested
in is that of the high energy density of states (at the sphere
level). We do so with an eventual goal of examining what aspects of
this high energy spectrum remain valid in ``little string
theory''. Note that since LST does not contain gravity, a thermal
equilibrium is well defined even though there is no weak coupling.

Since the discussion of qualitative features of free strings is
sufficient for us, we restrict our attention to the simplest model on
the worldsheet, i.e.  free bosons and fermions.  However, we allow for
an arbitrary central charge $\hat{c}_{eff}$\footnote{The reduced
central charge  is defined as usual, $\hat{c}=
\frac{2}{3}c$. We will refer to $\hat{c}$ as the central charge in the
following.}  (in light cone), and an unknown string tension
$M_{eff}^2$. We allow this freedom since the immediate information we
have about LST is the Hagedorn temperature which only determines the
combination
\beq
\label{crgtns}
{\hat{c}_{eff}\over M_{eff}^2}={4N\over M_s^2} \eeq where $N$ is the number
of NS 5-branes and $M_s$ is the spacetime string tension. This can be
derived from the Euclidean near extremal solution by measuring the
periodicity of the time direction, which turns out to be the same for
all values of the energy density. For example, the free string model of
LST in \cite{juan1} has strings of central charge 4 (4 bosonic
coordinates in light cone and their supersymmetry partners) and a
tension $M_s^2/N$.  However, the Matrix model (with a single unit of
null momentum) suggests a tension of $M_s$ and central charge
$4N$. Hence we would like to keep the central charge and tension
arbitrary (subject to the constraint (\ref{crgtns})), and see which
one better fits the ``data''.

There are several ensembles that one might use. One can use 
either the microcanonical or the canonical ensembles, and
then one can use either a compactified or uncompatified space. It is
at times stated that the canonical ensemble is unreliable in a theory
with a Hagedorn density because there are large fluctuations in
thermodynamic quantities. For example the fluctuations of the average
energy are:
\begin{equation}\label{flucen}
{<E^2>-<E>^2\over <E>^2}\sim 1 
\end{equation}
or much larger (as we remind the reader shortly).  However, the CGHS
black hole can be formulated just as well in Euclidean signature as in
Lorentzian one, therefore one expects the holographic relation to hold
just as well in the Euclidean/canonical ensemble case. In fact we will
show that the CGHS black hole predicts large fluctuations just below
the Hagedorn temperature. Henceforth we restrict our attention to
Euclidean signature and to the canonical ensemble.

One can also work either in compact on non-compact space, i.e., LST
either on $R^{5,1}$ or $R\times T^5$ where $T^5$ has some volume
$V$. We compute the thermodynamics in both cases and examine their
behavior as $T\rightarrow T_H$. In the case of compact space we will
also work in the limit $V\rightarrow \infty$, however we will always
take $T\rightarrow T_H$ first. Remember that these two limits do not
commute in a free string theory. The limit in which
$V\rightarrow\infty$ first for fixed $T$ converges to the
uncompactified limit, whereas when $T\rightarrow T_H$ first , for
fixed V, the result is different. One obtains a result which is
independent of $V$ and different from the non-compact case.

We will incorporate an arbitrary central charge by having 4 bosons and
fermions (in light cone) associated with target space coordinates of
the LST, and $D=\hat{c}_{eff}-4$ compact bosons and fermions, which model
''internal`` degrees of freedom, in the scenario of a large central
charge.  The spacetime directions will be taken to be either
non-compact or compact with a large volume. The compactification radia
of the internal CFT will remain fixed as we vary the temperature or
the space volume. To keep the Hagedorn temperature fixed, the
effective string tension scales as $M_{eff}^2 =
\frac{{M_s}^2}{\hat{c}_{eff}}$.

The mass formula for closed strings is the familiar: \bea M^2 &=&
2(n+n') + \vec{L}^2 +\vec{L'}^2 \nonumber \\ n'-n &=& \vec{L} \cdot
\vec{L'} \eea where we set the effective string tension to one for
now, and recover it later. The vectors $\vec{L}, \vec{L'}$ are momenta
and windings for the compact fields, rescaled to set the radia to
one. The Jacobian of this rescaling cancels in the expression for the
free energy.  The resulting Free energy is (approximating the discrete 
sum by an integral): 
\beq \beta F = V_5
\int d^5 p \int d\vec{L} d\vec{L'} \int dn \,\rho(n) \,\rho(n+ \vec{L}
\cdot \vec{L'}) \,e^{-\beta \sqrt{p^2 + M^2}} 
\eeq
with :
\beq
\rho(n) = \frac{1}{2^{(D+4)}} \, \frac{(D+4)^{(D+5)/4}}{n^{(D+7)/4}} 
\,e^{\pi\sqrt{n(D+4)}}
\eeq

Any potential divergence in the free energy near the Hagedorn
temperature comes from a saddle point where $n \rightarrow \infty$, so
we can expand the integrand in that limit. This gives: 
\beq 
\beta F =
V_5 \int d^5p \int dn \frac{1}{4^{(D+4)}} \,
\frac{(D+4)^{(D+5)/2}}{n^{(D+7)/2}} e^{2\pi\sqrt{n(D+4)}-
\beta\sqrt{p^2 + 4n}}\times
\eeq
$$\times \int d\vec{L} d\vec{L'}
e^{\frac{\pi\sqrt{D+4}}{2 \sqrt{n}} \vec{L} \cdot \vec{L'}} 
e^{{-\beta\over {4 \sqrt{n}}} (\vec{L} + \vec{L'})^2} $$
One can now perform the integrals over $\vec{L}, \vec{L'}$. Since
these integrals are not divergent as the temperature approaches the
Hagedorn temperature, one can substitute $\beta = \beta_H = \pi
\sqrt{D+4}$.

The expression for the free energy now becomes: 
\beq 
\beta F = V_5
\int d^5p \int dn \frac{1}{4^{(D+4)}} \,
\frac{(D+4)^{(D+5)/2}}{n^{(D+7)/2}} e^{2\pi\sqrt{n(D+4)}-\beta\sqrt{p^2 + 4n}}
\eeq
$$\times
\frac{1}{2^D}\left(\frac{8\sqrt{n}}{\sqrt{D+4}} \right)^D $$
which simplifies to the following expression (omitting D independent
numerical factors): \beq \beta F = V_5 \int d^5p \int dn \,(D+4)^{5/2}
\, n^{-7/2} \,e^{2\pi\sqrt{n(D+4)}} \,e^{-\beta\sqrt{p^2 + 4n}} \eeq

Now, define $m^2 = 4n$. We restore dimensions by using an effective
string mass $M_{eff}$, and use $D+4 = \hat{c}_{eff}$. This gives, again
up to numerical factors: \beq
\label{free}
\beta F = V_5 \int p^4 dp \int \frac{ M_{eff}^5 dm}{m^6}
{\hat{c}_{eff}}^{5/2} 
e^{\pi \sqrt{\hat{c}_{eff}}\frac{m}{m_{eff}}} 
e^{-\beta \sqrt{p^2 +m^2} }. \eeq 
We find that the free energy is extensive in $V_5$ and finite as
$\beta\rightarrow\beta_H$, and that these statements are independent
of the string model used (the precise value does depend on
$\hat{c}_{eff}$, which might eventually help determine $\hat{c}_{eff}$
for LST, if other obstacles, described in this paper, are overcome
first). Even though we calculated this behavior for a free worldsheet
we expect it to hold for any critical string at $g_s=0$ string.


The same calculations above can be repeated to find the free energy in
the case of a large compact spacetime directions. One finds the
qualitative behavior: \beq
\label{compact}
\beta F \sim \log( \beta- \beta_H) \eeq 
and independent of $V$. Again, this behavior is the same for the two
toy models that we are using, and we expect it to hold for every
$\sigma$-model.

As mentioned above, in the compact space case the behavior near the
Hagedorn temperature is very different from the non-compact case, and
in particular the free energy is no longer extensive. The difference
between the non-compact and compact case arises due to the dynamics of
long strings, which are allowed to wind on any compact
directions. Below we compare these results with LST thermodynamics and
find that even for a compact space the free energy is extensive. This
is a first hint towards the dynamics of LST - it is in a phase which
suppresses long strings.

One can now discuss other thermodynamics quantities, in the two
cases. In the non-compact model the energy stays finite near the
Hagedorn temperature \cite{aw}. In the compact case the energy
diverges as one approaches the Hagedorn temperature: \beq E =
\frac{1}{\beta- \beta_H} \eeq 

In addition energy fluctuations are large, and behave as \beq
{<E^2>-<E>^2\over <E>^2}\sim 1. \eeq The large fluctuations associated
with the canonical ensemble can also be attributed to the behavior of
the long strings. Near the Hagedorn temperature a finite fraction of
the energy is stored in a single string. This makes it increasingly
difficult to equilibrate the string gas as one approaches the
Hagedorn temperature.

Next, we would like to compare this behavior to the near-Hagedorn
behavior of LST. We do so by studying near extremal fivebrane
solution, to which we turn now.

\section{Near Extremal NS 5-branes}

We review here the gravity background holographically dual to LST at a
finite temperature. The zero temperature duality was discussed in
\cite{abks}. Roughly, as one approaches the boundary (UV) the solution
asymptotes (exponentially fast) to the weakly coupled throat region of
the CHS background \cite{chs}. At the interior (IR) there are several
interesting regimes, described by eleven dimensional supergravity,
which however will not concern us here.

At a finite temperature the solution used in \cite{abks} has to be
replaced by a near extremal version. Though the exact solution is not
known for all temperatures, it is becomes very simple in the limit that
the energy density becomes large, since then the entire solution
resides in the throat region and can be described using string theory.
The near extremal fivebrane solution was written down in
\cite{horstro,gidstro}. In terms of appropriately rescaled quantities
(after decoupling) the solution is (using the string frame):
\bea &ds^2 = dt^2 (1- \frac{\mu}{u^2}) +dx_i^2 + \frac{N}{u^2}(du^2
(1- \frac{\mu}{u^2})^{-1} +u^2 d\Omega_3) \nonumber \\ &e^{2\phi} =
\frac{N}{u^2} \eea where $u$ is related to  the radial
coordinate away from the brane, and $\mu$ is the energy density above
extremality. The index $i=1,...,5$ corresponds to directions along the
brane, and $d\Omega_3$ is the metric of the unit 3-sphere.

The near extremal solution can be brought into a more familiar form by
a redefinition of coordinates, $u = \sqrt{\mu}\cosh{r}$ (we also
rescale the coordinates along the brane): \bea
\label{cghsbh}
&ds^2 = N \left[ \,dt^2 \tanh^2(r) +dx_i^2 + dr^2 
+d\Omega_3 \right] \nonumber \\
&e^{2\phi} = \frac{N}{\mu \cosh^2(r)}
\eea

We identify the background to be made out of the two dimensional
black hole \cite{witten, cghs}, a level $N$ supersymmetric WZW and an
$R^5$ component\footnote{The extremal solution is a linear dilaton
direction times a WZW times $R^6$.}. We use the Euclidean version of
this black hole to describe the canonical ensemble. Accordingly, the
Euclidean time variable is taken to be compact, with periodicity
$2\pi$ in the conventions of (\ref{cghsbh}).

The geometry of the two dimensional black hole is that of a
semi-infinite cigar.  The asymptotic circle closes  at the tip,
which is located at $r=0$. The string loop expansion is controlled by
the value of the dilaton at the tip of the cigar:
\beq
{g^2_{s,tip}} = \frac{N}{\mu}
\eeq
We see therefore that the string loop expansion is expansion in
inverse energy density. This is the primary reason that string loop
corrections in this background can change qualitative features of the 
 thermodynamics.

It is worth noting that issues of decoupling are  not relevant
here \cite{sm, ms}. If the asymptotic string coupling
is small enough, although not strictly taken to zero, there will still
be a large range of energies in which the horizon of the black hole
will be in the throat region and our analysis will apply\footnote{The
distance from the tip of the cigar to the asymptotic region is
proportional to $\sqrt{N} log(g_{s,tip}/g_{s,asymp})$.}.

The tree level thermodynamics of the solution (\ref{cghsbh}) is well
known.  The system has a fixed Hawking temperature, $T_H \sim
\frac{M_s}{\sqrt{N}}$, regardless of the energy.  We calculate the
tree level free energy in the appendix and confirm that $F=0$ for each
value of the energy density $\mu$ (parameterized by the value of the
dilaton at the tip, $\phi_0 =\phi(r=0)$).

We find then that for fixed boundary conditions, there is a family of
solutions to the Euclidean equations of motion. These solutions have
degenerate action.  It would seem then that the parameter $\phi_0$ has
to be summed over, and represents a flat direction in the path
integral. Summing over $\phi_0$ would result in a divergence. However,
in \cite{ss} it was argued that the mode changing the parameter
$\phi_0$ is a non-normalizable mode, and should not be summed over in
the path integral.

The next step is to study the system at finite energy, namely study
string loop corrections\footnote{ Some aspects of the quantum corrections to the CGHS black hole 
were considered in \cite{stretched}.}. This is done in the next section.

\section{First Contributions to the Free Energy}

\subsection{Scalings of Correction Terms}

We discuss now the corrections to the leading order action, and their
effect on various thermodynamic quantities. Higher derivative terms in
the ten dimensional action were discussed in \cite{gw,r4}. They
include the famous $R^4$ term \cite{gw}, and many other terms,
related by supersymmetry. In order to organize the corrections to the
thermodynamics, we keep the number of fivebranes, $N$, large and
fixed, and arrange the possible corrections in powers of $N$.

We estimate therefore the scaling with $N$ of a general higher
derivative term, and then discuss the known terms in \cite{gw,r4}.
The metric as written above, equation (\ref{cghsbh}), is proportional,
in our notations, to $N$ (in the string frame).  Both $H$ and
$g_{str}^2$ are also proportional to $N$. Therefore, in the Einstein
frame one has: \bea &g_{..} \sim \mu^{1/4} N^{3/4} \nonumber \\
&H{...} \sim N \nonumber\\ &e^{\phi} \sim N^{1/2} \mu^{-1/2} \eea
where the dots indicate the position of spacetime indices.  As a check
of this scaling, all the terms in the leading order action \beq
\sqrt{g} \left[R + H^2 + (\partial \phi)^2 \right] \eeq scale
uniformly like $N^3\mu$.

We now can discuss the scaling of the correction terms.
The most general putative correction  term has the schematic form:
\beq
\label{term}
\sqrt{g} e^{b\phi} R_{....}^k {\partial_.}^l {H_{...}}^t {g^{..}}^p
\eeq

where both $b$ and $p$ can be negative or positive, and $l,k,t$ are positive.
We take negative $p$ to indicate metric with lower indices.

There are some constraints on the general term (\ref{term}). First,
the index structure of the above term demands: \beq 4k+l+3t = 2p \eeq
This guarantees that the term in the action is a scalar under general
coordinate transformations.

In addition, suppose we are interested in a particular power $s$ of the energy
$\mu$. The string loop expansion is controlled by the value of the dilaton 
in the tip of the cigar.  Therefore, the integer $s$ is related to
 the genus of the string diagram giving rise to the particular correction 
term (s= 1-g,where g is the genus).  For the general term above the power of
 $\mu$ is:
\beq
4s= 5-2b+k-p 
\eeq

We now can discuss the $N$ scaling of the general term (\ref{term}).
The power of $N$ of such a term is denoted by $A$, and is given by:
\beq
4A= 15 +2b +3k + 4t -3p
\eeq

In order to simplify this expression for $A$ 
 we can use the two relations  above to write $A$  in terms of the 
positive quantities $k,l,t$,  giving:
\beq
A = 5-s -k - \frac{l+t}{2}
\eeq

For a given genus, the power of $N$ is determined by $k,l,t$ , and the
two constraints above can be used to determine the needed $p,b$ to
complete the correction term (\ref{term}).

There are possible corrections to the action already at string tree level.
These $\alpha'$ correction  terms are all of $s=1$ form.  For example,  the 
$R^4$ term has $k=4$, and therefore scales like $\mu N^0$.  In fact, one 
can show that all 
the other leading order tree level corrections 
 scale like $\mu N^0$.

We are more interested in the  leading order corrections in the energy,
 coming from 1-loop terms in string theory. All such terms  have $s=0$.  For
 example, 
the  one-loop $R^4$ term has $k=4$, and therefore 
scales as $ \mu^0 N $. By a direct check, all other one loop terms scale 
similarly.  This   includes terms coming from up to 8 point scattering.

 To summarize, the leading corrections in one loop are all of the same
order, and scale like $\mu^0 N$. Compared to the leading order action
they are suppressed by $ \mu N^2 $.  We assume these corrections are
non-zero. In this case they represent the leading order correction to
the thermodynamics.

It is of some interest to calculate the coefficient of the
corrections. In addition to confirming that it is non-zero, its sign
has some significance in the thermodynamics. We elaborate on this
point below.

\subsection{Corrections to $E(T)$}

In the presence of corrections to the leading order action, the
solution (\ref{cghsbh}) deforms slightly.  The deformation is
controlled by the size of the corrections $\eta = \frac{1}{N^2
\mu}$. We are interested in the deformed solution asymptotically in
the $r$ direction, so we can observe a shift in the temperature.  One
has schematically: \beq I = I_0 + \eta I_1 + \cdots \eeq where $I_0$
is the leading order action (\ref{10d}), and $I_1$ are the leading
order corrections at one loop, discussed above.

 To keep track of the scaling of corrections to the thermodynamics, we
follow an outline of the calculation. An exact calculation would
require a detailed knowledge of the perturbed action $I_1$. Rather, we
are interested in extracting the dependence of the temperature
correction on the parameter $\eta$.

We are interested in the change in the asymptotic value of $G_{tt}=
h^2(r)$.  The equation of motion for this metric component reads: \beq
h''-2h'\phi' = \eta \frac{\delta I_1}{\delta G_{tt}} \eeq
We define the right hand side of this equation to be the source,
denoted by $J_1$.

Far enough from the tip, the background becomes approximately the
linear dilaton vacuum.  We define the small fluctuations: \bea &h = 1+
\delta h \nonumber \\ &\phi = -r +log2 + \phi_0+ \delta \phi \eea

Here we neglect the mixing with other small fluctuations. A complete
calculation would require, of course, diagonalizing the complete
matrix of quadratic fluctuations in this background. Clearly the
complete calculation retains the scaling with $\eta$ demonstrated
here.

One gets then the  following  equation for the small fluctuations:
\beq
\label{fluc}
\delta h '' + 2 \delta h' = J
\eeq

We note that the dilaton fluctuations drop out asymptotically, in the
linear dilaton regime.  The source in this equation $J=J_0+\eta J_1$
consists of two terms. The first term $J_0$ is independent of the
energy density, and is of no interest to us. The second term $J_1$ is
the one-loop correction defined above.

The solution to equation (\ref{fluc}) is a linear combination of the
two solutions of the homogeneous equations.  Therefore: \beq
\label{sol}
\delta h = a_1 + a_2 e^{-2r}
\eeq

The two a-priori independent coefficients are determined by demanding
regularity of the metric near the origin, as usual when determining
the temperature.  We are interested in the coefficient of the constant
solution, $a_1$, which affects the asymptotic radius, and hence the
temperature.

The coefficient $a_1$ is given by an overlap integral of the constant
solution with the source $J$.  The only dependence of $a_1$ on $N,
\mu$ appears explicitly in the parameter $\eta$.  Therefore $a_1$ can
be written as a series expansion in the parameter $\eta$.  The $\eta$
independent part only renormalizes the value of the Hagedorn
temperature.  The leading dependence on the energy density $\mu$ comes
through: \beq \delta h(r) \sim \eta \qquad\mbox{as} \,\, r \rightarrow
\infty \eeq

The shift in $h(r)$ gives the following correction to the inverse
temperature\footnote{There is also a correction to the relation
between the parameter $\mu$ and the energy density, but it has a
subleading effect in the string loop expansion.}.  \beq
\frac{\beta-\beta_H}{\beta_H} \sim \eta \eeq

The coefficient of the correction becomes significant at this point.
If the proportionality constant in the last equation is positive, the
system approaches the Hagedorn temperature from below, at high
energies. Asymptotically the system has a positive specific
heat. Therefore the canonical ensemble is well-defined, though it has
large fluctuations as explained above.

The scenario of a negative coefficient is very different: the system
approaches the Hagedorn temperature from above, and has a negative
specific heat. The canonical ensemble does not exist for high
energies. We assume this is not the case and the coefficient is
positive.

This gives the following modified relation between the temperature and
the energy density: \beq \mu \sim \frac{N^{-5/2} {M_s}^5}{\beta-
\beta_H} \eeq where ${M_s}^2$ is the spacetime string tension. Every
string model will have to reproduce this expression, including the
scaling with $N$.

\subsection{Corrections to the Partition Function}

The modified energy-temperature relation determines the thermodynamics
of the system. Such a relation is expected to arise from an effective
action of the form: \beq \beta F \sim log (\beta- \beta_H) \sim log
(\mu) \eeq

This kind of contribution comes from a dependence of the effective action
$I= \beta F$ on the logarithm of $g_{str}$, the string coupling at the tip.
 Since this dependence might seem unexpected, we demonstrate 
how it can be generated by the one loop perturbation.

The leading perturbation to the effective action is a sum of two
 terms. The first comes from the substituting the deformed solution in
 the original action, $I_0$. The second term comes from the action
 $\eta I_1$, evaluated on the unperturbed solution.  Logarithmic
 contribution can arise from terms in the integrand that are constant
 in the linear dilaton regime. Since the length of the linear dilaton
 regime is proportional the $\log(g_{str})$, as shown above, this
 gives the desired effect.

Terms in the integrand which are constant in the linear dilaton regime are
 quite natural. We demonstrate here some possible sources for them.
For example, the deformed solution for the metric component
$G_{tt}$ is
given by (\ref{sol}). This gives the following increment to the 
tree level effective action:
\beq
\delta I = \beta \int dr   e^{2r}\left[-2 \delta h''+ 4 \delta h + 
\cdots \right]
\eeq

The integrand in this expression contains a constant term, where the
 decay of the fluctuation $\delta h $ is compensated for by the
 prefactor. There are also divergent terms, proportional to the
 temperature shift. This terms are regulated by considering
 differences in the action, as demonstrated in evaluating the leading
 term above.  This procedure can also give rise to constant terms.

 Another set of contributions to the free energy comes from
a direct substitution of the original solution into the new action.
 These contributions  can also contain constant terms in the integrand.  
To demonstrate
this we need to consider the general correction term, evaluated in the
linear dilaton background.  In this background the only dependence on
the radial coordinate is through the dilaton profile. Furthermore,
derivatives of the dilaton are constant. The only dependence on the
radial coordinate comes from the exponential of the dilaton. One loop
terms come with the dilaton exponential raised to zeroth power. We
find therefore that the corrections terms, evaluated in the linear
dilaton regime, are all constant, and give rise to the desired
logarithmic dependence.

Therefore, logarithmic behavior of the free energy is natural in the 
linear dilaton throat.
   This logarithmic behavior of  leads to large fluctuations in the energy.
 One gets, as was discussed above:
\beq
{<E^2>-<E>^2\over <E>^2}\sim 1
\eeq

The coefficient here is of order $N^0$. We elaborate on the issue of
large fluctuations below.

In a critical string theory, the one loop contribution results in
effects that are quite different from the ones found above. At
temperatures near the Hagedorn temperature there is a mode, winding
around the Euclidean time circle, which becomes light.  The effective
description of this mode is discussed in \cite{aw}.  The situation is
similar to the one described by Rohm \cite{rohm}, when considering
Sherk-Schwarz supersymmetry breaking in string theory. The breaking by
the boundary conditions on the circle results in a dilaton tadpole,
and a one loop cosmological constant. The modified equations of motion
have no longer a static solution.

We now argue that the effects discussed in \cite{aw, rohm} are
subleading to the ones discussed above. The effects of SUSY breaking
by the boundary conditions on the Euclidean time circle can be
computed in supergravity, as the system is at a low temperature
compared to the string scale. The induced cosmological constant is of
order $(\frac{\beta}{l_p})^8$ compared to the leading order action. In
the present background this gives a suppression by $N^3 \mu$. To
compare, the effects discussed above are suppressed by $N^2 \mu$, as
compared to the leading order action.

\section{Another Look at Large Fluctuations}

The expression for the free energy, or the energy as a function of the
temperature, reveals a problem in using the canonical ensemble when
studying the thermodynamics of the system.  The statistical
fluctuations in the energy density are found to be: \beq (\Delta
\mu)^2 \sim N^0 \mu^2 \eeq

 As one approaches the Hagedorn temperature from below, (normalized)
fluctuations in the energy remain finite. This makes the canonical
ensemble not very useful when calculating thermodynamic
averages. Physically, this stems from the attempt to keep a system
with an exponential density of states in an equilibrium with a heat
bath.

However, the canonical ensemble is still holographically dual to the
Euclidean black hole, and it is useful to study these large
fluctuations in the holographic description.  At first sight, one
might expect that the mode that changes the energy of the system
becomes nearly massless at high energies, leading to the
aforementioned fluctuations. However, as mentioned above,this mode is
found to be a non-fluctuating mode \cite{ss}.  We present here an
alternative picture for the large uncertainty of the energy. A related
picture has appeared before in \cite{stretched}.

The system has a large number of normalizable, fluctuating
modes. Their specturm was written in \cite{dvv}, and the behavior in
the linear dilaton regime was studied in \cite{ms}.  The action for
small fluctuations around a linear dilaton background is: \beq I =
\int dr e^{2r} \left[ G^{ij} \partial_i \psi \partial _j \psi \right]
\eeq

Here $G_{ij}$ is the string frame metric, and $\psi$ is a typical
fluctuation, for example a fluctuation in the metric polarized along
the brane directions.  We study the radial profile of the
fluctuations. Our radial coordinate is related to \cite{ms} as $z=
\sqrt{N} r$. We also set the string tension to one for now.  The modes
$\psi$ were normalized to absorb any additional factor in the action.

 The  modes of the action
are parameterized by $s = N \omega^2$. Their asymptotic behavior is:
\beq
\psi_{\pm}(\omega) = \exp( \beta_{\pm}(\omega) r - i\omega t )
\eeq
with
\beq
\beta_{\pm} = -1 \pm \sqrt{1-  N \omega^2} 
\eeq

For energies above the gap $\omega_0 = \frac{1}{\sqrt{N}}$, the modes
are normalizable. Their fluctuation scale is set by their action. We
now proceed to estimate the fluctuations of a single normalizable
mode.

Assume we are working with a finite cutoff, such that the length of
the tube is finite and equals $L$. Imposing boundary conditions at
$r=L$, and regularity conditions at the horizon, results in a discrete
spectrum.  Generically, this spectrum has spacing in energy $\omega$
in the order of $\frac{1}{\sqrt{N}L}$.  This assumption is invalidated
if the regularity at the horizon sets a complicated, $\omega$
dependent, relation between the two solutions $\phi_{\pm}$. We assume
this is not the case for generic normalizable modes.

We assume therefore that the boundary conditions pick a solution of the 
schematic form:
\beq
\psi(\omega) =   e^{-r}\sin( \sqrt{N} \omega r) \sin(\omega t)
\eeq
The exact form of the trigonometric functions is immaterial.

 For large energy modes the action is:
\beq
I(\omega) = \int dr \,\omega^2 \sin^2 (\sqrt{N} \omega r) \sim L \omega^2
\eeq

The last step comes from averaging a rapid fluctuation over the length
$L$, which is much larger than the period. We find therefore that with
a finite cutoff $L$ , the typical fluctuations of the mode
$\psi(\omega)$ are given by: \beq (\Delta \psi)^2 = \frac{1}{L
\omega^2} \eeq

Each mode $\psi(\omega)$ couples to the mode which moves the tip of
the cigar by a certain form factor. This coupling is computed at tree
level , and is therefore a general function of the form $G(s)
=G(N\omega^2)$. The fluctuations of the location of the tip, resulting
from the fluctuations in all normalizable modes, are estimated to be:
\beq (\Delta a)^2 = L \sqrt{N} \int d\omega \, G(N\omega^2) \frac{1}{L
\omega^2} \eeq where we denote the radial position of the tip of the
cigar (as defined by the zero of the metric component $G_{tt}$) by $a$, and its variance by $(\Delta a)^2$.  The normalization
factor is needed when converting a discrete sum with spacing
$\frac{1}{L \sqrt{N}}$ into an integral. We note that the dependence
on the cutoff $L$ drops off, as it should.

The form factor $G(s)$ is unknown, but is expected to falloff at large
frequencies, in order to yield a finite result. Impose a large
frequency cutoff $\Lambda$, and rescale $k= \omega \sqrt{N}$. This
gives: \beq (\Delta a)^2 = N \int^{\Lambda \sqrt{N}} \frac{dk}{k^2}
G(k^2) \eeq Under the assumption of convergence of the integral, the
$N$ dependence disappears from the integral. The fluctuations in the
location of the tip of the cigar are of order $\sqrt{N}$, in proper
distance. Therefore, in our notations, fluctuations in the coordinate
$r$ are of order one.

This translated, via the exponential relation between the radial
location of the tip and the energy density, to the following: \beq
(\Delta \mu)^2 \sim N^0 \mu^2 \eeq

We see therefore that under reasonable assumptions about the behavior
of the normalizable modes, their collective effects result in large
variance of the energy density. This fits with the expectations of the
dual configuration, namely the canonical ensemble of LST.

\section{The Coiled Phase of Strings}

\subsection{The Discrepancies}

We would now like to compare more explicitly the free string model and
the results obtained from the black hole. The free string results are
either:
\beq
T\rightarrow T_H,\ V\ arbitrary, \ \ \ <E> = {1\over {\beta-\beta_H}} 
\eeq
$$T\rightarrow T_H,\ V=\infty \ \ \ \ {<E>\over V}= finite.$$
Again, these relations are expected to be robust in critical string
theory. Both relations are expected to be generic in the $g_s=0$
string.

On the other hand in ``little string theory'' the relation we obtained is:
\beq
T\rightarrow T_H,\ V\ arbitrary, \ \ \ {<E>\over V}={1\over \beta-\beta_H}
\eeq
(and we have neglected factors of $N$, and dimensions are corrected
using $M_s$).
We clearly see that the free string model misses qualitative features
of the thermodynamics. Next we would like to see whether there is some
natural modification of the string dynamics that will go towards
explaining the near-extremal thermodynamics.

Let us examine the non-compact partition function to see what
modification can give us the correct result. Schematically the
non-compact space free energy from section 2 is
(after integrating over the momenta) 
\beq 
\beta F \sim V_5\int {dm\over m^{7/2}} e^{-(\beta-\beta_c)m}.
\eeq

We would like to examine what modification to the partition function
can change  the power $m^{-7/2}$ into $m^{-1}$. Of course, there is
no unique answer, but for later use we poit out that a possible   modification
is changing the energy by a $ln(m)$ term. More precisely the correct
mass of a state is given by \beq m_{correct}=m-{5\over2\beta_c}ln(m)
\eeq In this case, we will obtain the correct behavior for the
partition function. 

The general motivation will be following. The mass of the string is
roughly its length. We are therefore interested in a string
interaction that reduces the energy of the string by $ln(l)$. We will
see that there are such natural interactions, and we will focus on a
self-intersection interaction which does that. Because the string now
self-attracts it prefers to be coiled rather then large, solving the
problem of dominance of long strings, and suggesting that we should
consider a new phase made out of coiled strings.

\subsection{General Considerations}

We have seen that it is difficult to understand the near Hagedorn,
i.e., high string excitation, behavior of LST in terms of a familiar
string theory. In this section we would like to propose a different
picture which, although we can not make precise, seems to remedy the
situation. The upshot will be a new phase of strongly interacting
strings with new qualitative features\footnote{Another possibility is
the existence of open strings in the system,  this was suggested to   us
by O. Aharony. The thermodynamics of open string sectors was recently
discussed in \cite{rab}.}.

Let us begin by highlighting two main properties that such a solution
should have:

\begin{itemize}

\item First of all it should suppress long strings: As we have seen
above, when space is compactified on a large torus, the contribution
of long winding strings makes the free energy non-extensive. In order
to obtain an extensive answer, we would like to suppress the
contribution of these strings to $S(E)$.

Actually, this aspect is closely related to the non-gravitational
nature of the theory.  As emphasized by Susskind \cite{susssize}, one
generally expects that in gravitational theories, objects that naively
contribute many states to the entropy become large, such that their
contribution does not violate the Bekenstein bound (For example, the
ground state of the string grows when taking into account more and
more oscillators \cite{lenigor}). Hence it is natural in critical
string theory for highly excited strings to tend to be very large.

 In our case, since there is no gravity and there are off-shell
observables, we expect the string to prefer to be coiled rather then
long  at large excitation number.
This observation is  model-independent, and relies only 
general  aspects
of the relation between the number of degrees of freedom in a large
volume and the size of highly excited objects.

\item Secondly, the modification should have some reasonable
space-time interpretation. Suppose we have a long string, or two long
strings, then a reasonable space-time interpretation requires that
when pieces of the strings are far away then the force between them
will fall off at least as fast as  the exchange of massless particles.
\end{itemize}

 The conclusion is that we would like to modify the
string models that we used before, which were some CFT on a
worldsheet, by a strong attractive interaction. The new string is
such that it can not be written in terms of a local worldsheet action
(otherwise, we do not expect to  evade the problems outlined above).

As an extreme, we can try and model the string with purely local
interactions in space-time. In this case it would be a
self-intersection interaction. We will argue that this is indeed what
happens in the light cone frame. There are various kinds of
self-intersection interactions that one can write down, but one
expects that the most relevant attractive self-intersection
interaction is simply such that one looses a fixed amount of
energy for every self-intersection (in some regulated version of the
worldsheet).

This is, of course, not precise
for the LST. For example, two segments of strings that intersect at a
very small angle such that they are almost parallel are almost BPS and
hence there is no force between them, whereas anti-parallel strings
attract each other. Hence the interaction is not as simple as we suggest. 
Note, however, that the average over these
configurations certainly gives an average attractive interaction.

Two more comments are due. First, we have emphasized the effects of
self-intersections but clearly if there is a strong intersection
interaction, then each string in the thermal state interacts strongly
with the background. How can we take into account this interaction ?
We do not have the complete answer to this question, but we can
suggest the following observation. One expects that if one uses a
``mean field approximation'' in which one computes the single string
state statistics in a homogeneous background (encoded in the values of
some order parameters), then the contribution to the energy of the
string will be proportional to the length of the string. This means
that the ``mean field approximation'' parameters of the background
renormalize the string tension, but may not correct other terms. In
particular the self-intersection term (which we will show gives the
logarithmic correction in the exponent) is left uncorrected. The
string tension may indeed be renormalized, but the Hagedorn
temperature already measures the physical tension, after this
renormalization has been taken into account (actually, a self
intersection interaction also renormalizes the string tension, and the
same argument applies to this renormalization as well).

Finally, in our discussion of the interaction we have used some kind
of intuition about locality. This might be dangerous in a theory with
T-duality in which space-time is not a well defined object - do we
mean the initial, say, torus or its T-dual ? In our discussion what we
mean is that in the large volume limit, there is an interaction with the
properties discussed above, and in particular with approximate
locality in the large space-time. There are many other non-local
interactions, coming say from winding strings, but they decay as
$V\rightarrow \infty$. This is true in critical string theory, and when we turn
our attention to motivations Matrix description next, we will see that
the behavior is very similar in LST.

\subsection{Matrix model considerations}

It is instructive to check whether this picture is consistent with the
DLCQ description of LST \cite{abkss,higgs}. We are motivated by the
fact that counting the single string degeneracy is simplest in light cone,
and hence we would like to use Matrix theory to analyze it.

The Matrix model for LST on $R^{5,1}$ is the D1-D5 system which is a
1+1 dimensional sigma model on the ADHM moduli space. The latter is
parameterized by two integers: $N_0$ which is the number of instantons
and $N$ which is the rank of the $U(N)$ gauge group. In the Matrix
interpretation $N_0$ measures the null momentum and $N$ the number of
NS 5-branes. The Matrix model for the LST on $R^{1,1}\times T^4$ is
the same D1-D5 system, where the D5 is now compactified on $T^4$, and
we will focus on this model in our discussion. This model is a sigma
model on a target space which is a deformation of
${T^4}^{N_0N}/S_{N_0N}$ where $S_l$ stands for the group of
permutations of $l$ elements. The model is a deformation in the sense
of not being at the solvable orbifold point. Rather, it is at a
singular point in the moduli space of CFT. This point is the analogue
of the $\theta=0$ point of $R^4/Z_2$ \cite{aspin}. Because of these
singularities it is not clear how to analyze the model, but
fortunately these singularities are rather mild as far as the current
question is concerned. The reader is however warned that the analysis
  we present is rather speculative, but at this point we would like
primarily to check consistency with the picture put forward above. The
analysis also implies a new way of analyzing the D1-D5 system, and it
will be interesting to explore it more rigorously further.

We would like to analyze the spectrum relevant in the Matrix framework
(i.e., the correct energy scaling) and see whether it fits the line of
thought explained in the previous subsection. If the model was the
symmetric product then the analysis would have been straightforward. We
could go to a picture of ``long strings'' by going to the twisted
sector of the $S_{N_0}$ part of the symmetric group and obtain a
string with the correct scaling of energy, tension $M_s$ and central
charge $4N$ (or we could go to even longer strings with by using the
remaining $S_N$ symmetry to a string with central charge 4, but we
will not need this stage here). This would give us a Hagedorn density,
but we saw before that it does not reproduce the thermodynamics
correctly. This is not surprising since the CFT is not at the
orbifold point. 

Nevertheless, we would like to argue that such long strings are a good
starting point. The reason for that is that the entropy of strings far
away from the singularity is much larger then the entropy of strings
at the singularity. The reason is that the total central charge of the
CFT is $4N_0N$, whereas the effective strings at the singularity have
a much smaller central charge. This can be extracted from \cite{ab} in
which the states at the singularities were described in terms of a 1+1
field theory written in terms of a $U(N_0)$ vector multiplet. The
states at the singularity correspond to excitations along the flat
directions of this non-abelian gauge theory, and hence their central
charge is much smaller.

Hence, most of the states can be thought of as bulk states, i.e.,
these states for the most part are traveling away from the
singularities of the sigma-model and are unaware of the
singularity. In this sector one can try and go to long strings first
and then consider the effect of the interaction. These long strings
are the strings in spacetime and the interaction is precisely an
interaction which is localized when the string intersects itself -
i.e., it is an interaction of the same type that we were advocating
above.

Hence the Matrix theory description lends support to the proposal that
the modification is a strong self-intersection interaction. This
analysis also suggests a new way of analyzing the D1-D5 system.

\subsection{A Random Walk Model}

Let us now try and estimate how a self-intersection interaction in
light cone can change the partition function of a string theory. We
will not do so precisely but rather point to a  relation between this
model and a certain model  of self-attractive random
walks.  We will perform most of the computations in the latter
model.

 Modelling  dynamics  by a random walk
is familiar from polymer physics \cite{degennes}. In the context of
string theory Horowitz and Polchinski \cite{horpol} analyzed
corrections to the size of an excited string state using such methods
in the context of the black-hole/excited string correspondence
principle \footnote{There is, however, a difference in that there if
the string moves in d+1 dimensions, then the random walk is in d
dimensions. In our case we are motivated by the lightcone picture to
use a local intersection interaction for the 4 transverse coordinates
in light cone.}. Although most of the analysis there is not directly
relevant to our case \footnote{An important part of interaction there
is a long-range  gravitational self attraction, which we do not have
 here. Also,
it is not at all clear what might be the interpretation of the thermal
scalar in our context.}, it is worth noting that they also find
significant effects which lead to a  contraction of  string states.

Also, we will not be working precisely in the DLCQ of LST but rather in
a simplified model in which there are 4 transverse (to the lightcone)
coordinates and a simple self intersection interaction term, i.e., the
energy decreases by some fixed amount when the string self-intersects
in these coordinates. Other then this interaction term, one usually
takes the action for the string to be its area. We will take a
simplified model in which the weight of the string in the partition
function is its length. This is so because we are counting physical
states in lightcone, and we expect some equipartition between the
kinetic and potential term. Hence there would still be a linear
relation between the average size of the string as it fluctuates in
some quantum state and the energy of that state (after all in lightcone
 the theory is a collection of oscillators). In any  case, the
skeptic reader can view the following analysis as a toy model which on
the one hand reproduces some aspects of the partition function, and on
the other hand is convenient for the analysis of the self-intersection
interaction.
 
Under these assumption the partition function, without self
intersections, at some inverse temperature $\beta$ will be \beq \int
dl f(l) e^{\beta_c l -\beta l} \eeq where $l$ is the length of the
string (f(l) is a function which determined the degeneracy for a  given
length $l$). In relation to the usual quantization of strings we see that
in order to match to the usual free oscillator  picture we need to
assume that at leading order $l\sim \sqrt{n}$.

We would like to regulate this partition function in a way that
captures its interpretation as strings in space-time. The way to do so
is to approximate it as a sum over random walks, which will also
enable us to more precisely define what we mean by the number of
self-intersections. The length $l$ will now become a discrete variable
and the partition function is a sum over all random walks with $l$
steps with weights $e^{-\beta l}$. We will also assume the simplest
form of a random walk, i.e., a cubic lattice with nearest neighbors
jumps with equal probabilities.

 Adding now the attractive interaction, the partition function  becomes
\beq \sum_{random\ walks} e^{-\beta (l - g_{rw} J)} \eeq where $J$ is the
number of self-intersection \beq J=\sum_{i,j=1,..,l}
\delta_{w(i),w(j)}, \eeq  and $w(i)$ denotes the location of the
random walk at time $i$. $g_{rw}$ is some definite number which is
determined by micro-physics and therefore we can not estimate. 

 We come now to the main purpose of the random walk
model. Evaluating the number of self intersections for a simple random
walk of length $l$ and regarding it as a correction to the energy of
the random walk, we will obtain a logarithmic correction to the
energy. Of course, without knowing the coefficient $g_{rw}$ above we
will not be able check whether the logarithm in the exponent gives
eventually the right power of $l$ needed to correct the degeneracy of
the states, but it is interesting to get a logarithmic correction at
all. The reason is that in any higher dimensions there are no
logarithmic terms in the self-intersection number. Hence 4 coordinates
in lightcone is the maximum number for which one would expect to see this
 type of
universality class of strings. Fortunately this is precisely the maximal
dimension of LST.

Finally let us explain how the logarithm comes about. We would like to
evaluate $<J>_l$ on a closed loop of length $l$. We will use a Feynman
diagram like technique (combinatoric computations as well as
diagramatic techniques are described  in \cite{bryslade}. An
explicit formula for self-intersection number, for  open random walks in various dimensions, 
appears in  Brydges and Slade \cite{bryslade}). Since we are
evaluating a single insertion of $J$, the dominant contributions are
random walks in which start at point 0, propagate to some point X
after $i$ steps, return to that point after additional $j$ steps and
then return to the origin after additional $l-i-j$ steps. This
diagrams contributes (The factor $l^2$ in front is due to normalizing
by the probability that the random walk will return to the initial
point - i.e., will be closed): \beq <J>_l \, \propto l^2\int_{i+j<l} di dj dx
e^{-{x_1^2\over i}}{1\over i^2} {1\over j^2} e^{-{x^2\over(l-i-j)}}
{1\over {(l-j-i)}^2} \eeq where we have made the discrete sum over
steps into an integral but require $i,j>1$ (we are  not careful
with numerical coefficients). It is straightforward to evaluate this
integral and which yields in the large $l$ limit: \beq <J>_l \,\propto l +2
ln(l)+O(1)
\eeq The first term renormalized the string tension, and  the second term
is precisely what we wanted above - a change in the energy of the
string that is proportional to the logarithm of its space-time length.

\section{Conclusions}

In this paper we discussed the thermodynamics of ``little string theory''
at high energies. Studying string loop corrections to the holographic
dual we were able to go slightly off the Hagedorn temperature and
probe the physics as we approach that critical temperature.

The main surprise is that this physics is inconsistent with a picture
of free closed strings at high energies. The discrepancies are
fundamental and generic to all free ($g_s=0$) string models. This
provides a possible clue to the dynamics of the ``little strings'' at
high energies.

The main discrepancy has to do with extensivity of the thermodynamics
when the system is put at a finite volume. Free strings do not give
extensive results, due to the dominance of long strings at high
energies. The holographic duality teaches us that the ``little
strings'' do give an extensive result. This is a first hint towards
understanding dynamics in ``little string theory'': the ``little
strings'' do not form a single long string at high energies, rather
they prefer to clump in small coils.

We argue for a simple model which reproduces this behavior. This model
involves strings (in the lightcone) which interact only when they
self-intersect. This suggestion is motivated by several
observations. First, it seems that the interactions of the spacetime
strings as seen in the DLCQ description yields an effective
interaction of this form. Second, it is consistent with a local (on
large scale) dynamics in spacetime. In addition, the interaction is
different enough from the free string picture, so that a different
qualitative behavior is possible.

Indeed, a simple analogous random walk model shows significant changes
when an attractive self-intersection interaction is added. The changes
look qualitatively similar to what we need: strings tend to become
``coiled'', and the partition function gets corrections similar to the
ones needed to reproduce the thermodynamics of ``little strings theory''.

Clearly, more work is needed to establish this claim. A direct
relation to a random walk model is needed if one is to reproduce
exact, quantitative features of ``little string theory''. Perhaps
working in the DLCQ description, this relation can be clarified. We
hope to return to these issues in the near future.

\section*{Appendix: Vanishing of the Leading Order Free Energy}

We calculate here the leading order thermodynamic quantities. We use
the canonical ensemble, which corresponds to a Euclidean black hole
configuration with a compact time direction.  The basic quantity to
calculate is the free energy, which is obtained from the value of the
action on-shell \cite{gh}.

We start with the ten dimensional Einstein frame action:
\beq
\label{10d}
I =  \frac{1}{16\pi G_{10}}  \left[\int_M d^{10}x \,\sqrt{g} (R - \frac{1}{2} 
(\partial_\mu \phi)^2 - \frac{1}{12} e^{-\phi} H^2) + 2\int_{\partial M } K
\right]
\eeq

To fix the ambiguity in the total action we need a prescription for
boundary terms in (\ref{10d}).  Following \cite{gh}, we choose the
total action such that fixing the values of the metric at infinity,
but not its normal derivatives, is allowed.  The standard boundary
term, written above, involves the trace of the second fundamental form
of the boundary.

  We now transform to string frame metric $G$ by a conformal
transformation,  $g_{..} = e^{-\phi/2} G_{..} $,
where $H_{...}$ remains unchanged in the transformation. Also, to
compare to  notations  in \cite{witten}, we write $\Phi = -2 \phi$

In addition to the standard string frame bulk action, there is an
additional boundary term, which is, for the sphere at infinity: \beq
-(9/4)\sqrt{G}e^{\Phi} G^{rr} \partial_r \Phi \eeq where $r$ is the
radial direction.

Now we  perform dimensional reduction. Denote by $V_5$ the volume of the 
noncompact 5 dimensions, and by $V_{sph}$ the volume of the 3-sphere.
 We choose the following ansatz for the fields:
\bea
&ds^2 =  \left( {ds_2}^2 + {dx_i}^2 + N d\Omega_3\right) \nonumber \\
&\Phi = \Phi(r)  \nonumber\\
&H =  N dV 
\eea

Here $i=1,...,5$ are flat directions along the brane, and $d\Omega_3$
denotes the standard metric on $S^3$. $dV$ is the volume element on
the 3-sphere. This ansatz covers both the extremal and non-extremal
fivebrane solution, therefore it is suitable when calculating the
effective action.

One then gets 
\beq 
\label{action}
I = \frac{V_5 V_{sph}}{16\pi {l_s}^8} \int d^2x
\sqrt{G} e^\Phi \left[ R + G^{\mu\nu}(\partial_\mu \Phi)(\partial_\nu
\Phi) + 2/N \right] 
\eeq 

 The prefactor is normalized in
\cite{witten} to be 1, by a shift of the dilaton $\Phi$. We set the
prefactor to 1, and recover it later.

 We work in the ansatz for the two dimensional metric: \beq {ds_2}^2 =
N \left( dr^2 +h^2(r) dt^2 \right) \eeq Here $t$ is Euclidean time,
compactified with a period $\beta l_s$, where $\beta$ is
dimensionless. The physical temperature is then $\frac{1}{\sqrt{N}
\beta l_s}$.

 The action should include also boundary terms discussed above.
 The boundary terms in the ten dimensional string action, written in the 
present ansatz, give the following terms:
\beq
\label{boundary}
I_1 + I_2 =  \frac{2}{N} \beta e^{\Phi}h'
\eeq

Both are to be evaluated at the boundary at infinity only, where the
Dirichlet boundary conditions have to be imposed.

The bulk two dimensional action (\ref{action}) depends on second
derivatives of the function $h$. In order to find the equations of
motion we integrate by parts. This results in a boundary term we
denote $I_3$: \beq I_3= \beta \sqrt{G} e^{\Phi} \left[G^{\mu\nu}
\Gamma ^r_{\mu\nu} -\Gamma^{\mu}_{r\mu} \right] \eeq

The bulk action is then:
\beq
I = \beta \int dr e^{\Phi} \left[ 2h' \Phi' + 4h + h{\Phi'}^2 \right]
\eeq
 where prime denoted derivative with respect to $r$.

The value of the action on shell is also a total derivative. This is
easily shown using the equations of motion of the field $\Phi$.
Therefore the two dimensional bulk action (\ref{action}), when
evaluated on shell, equals the following terms, evaluated at the
boundary: \beq
\label{bulk}
I_3 + I_4 = \frac{1}{N} \beta e^{\Phi}  h \Phi'
\eeq

The boundary here includes the black hole horizon as well, due to the
different origin of this boundary term. This is still to be regarded
as a bulk action, though for this particular solution it reduces to
boundary terms.

 It is easy to show that the following solves the equations of motion:
\bea
\label{bh}
&h(r) = \tanh(r) \nonumber\\
&\Phi(r) = 2log\cosh(r) +a
\eea

Where $a$ is an arbitrary constant. This is the near extremal solution
(\ref{cghsbh}). The action of this solution should be compared with
the vacuum configuration: \bea
\label{linear}
&h(r) =1 \nonumber\\
&\Phi = 2r +a - 2log2
\eea
 
The additive constant in $r$ is chosen so that the field $\Phi$ has
the same asymptotic behavior.

The effective action of the black hole configuration is found by a
direct substitution in the above expressions.  On general grounds one
gets: \beq I_{total} = \beta F = \beta E - S \eeq

Here $F$ is the free energy of the configuration, $E$ is the average
energy and $S$ is the entropy. The first term $\beta E$ comes from
boundary contribution, and the second one from a bulk contribution.

The bulk contribution comes from equation (\ref{bulk}). This leads to
a divergent result for the black hole (\ref{bh}), but has to be
compared to the action of the linear dilaton vacuum. This gives: \beq
I_3 + I_4 = -\frac{1}{N} \frac{V_5 V_{sph}}{8\pi {l_s}^8} \beta e^a
\eeq where we have restored the prefactor set to 1 above.

The terms that get a contribution from the boundary at infinity are
given in (\ref{boundary}). They give the action: \beq I_1 +I_2 =
\frac{1}{N} \frac{V_5 V_{sph}}{8\pi {l_s}^8}\beta e^a \eeq

Combining the bove gives $F=0$ for every $a$. Also one can read off
the energy density and the entropy: \bea & \mu = \frac{E}{V_5} \sim
\frac{N}{g^2} \frac{1}{{l_s}^6} \nonumber \\ &S= \beta_H E \qquad
\mbox{with} \,\, \,\,\beta_H = l_s \sqrt{N} \eea This gives the
expected Hagedorn behavior, and the correct relation between the
energy density and the string coupling, as obtained by other methods.

\section{Acknowledgments}

We ate happy to thank O. Aharony, M. Aizenman, C. Bachas, T. Banks, M. Douglas,
J. Harvey, G. Horowitz, I. Klebanov, A. Lawrence, E. Martinec, G. Moore,
B. Pioline, A. Rajaraman, S. Shenker, L. Susskind, E. Verlinde and
H. verlinde for useful discussions.  MR would like to thank the theory 
group at the University of Chicago for hospitality while this work was
 being completed. The work of MB is supported by NSF
grant 98-02484. The work of MR is supported by DOE grant
DOE-FG02-96ER40959.

\end{document}